\documentclass[aps,pre,groupedaddress,showpacs,preprint]{revtex4}
\usepackage{graphicx,epsf}
\usepackage{xcolor}
\newcommand{\vvec}[1]{\mathbf{#1}}
\newcommand{\mathbb}[1]{\mathbf{#1}}

\usepackage[english]{babel}
\begin{document}
\title{Universal size and shape ratios for arms in star-branched polymers: theory and mesoscopic simulations}

\author{Ostap Kalyuzhnyi}
\author{Khristine Haidukivska*}
\author{Viktoria Blavatska}
\author{Jaroslav Ilnytskyi}

\affiliation{Institute for Condensed Matter Physics of the National Academy of Sciences of Ukraine, 1, Svientsitskii Str., 79011 Lviv, Ukraine, E-mail:wja4eslawa@gmail.com
}

\begin{abstract}

Star polymer undergoes a transformation from a group of loosely coupled chains at low number of arms to a dense hairy colloid at their high number. This change affects solubility, aggregation and rheological behavior and is of much practical interest. We study the range of size and shape properties of the star molecule and of its individual arms upon this transformation. Theoretical calculations are based on a continuous chain model and are performed in the first order in $\epsilon=4-d$. Computer simulations are done by the dissipative particle dynamics and in part by Monte Carlo method and the results demonstrate very good agreement with the selected set of properties known from the previous simulations. Theory and simulations provide qualitatively similar trends upon increase of the number of arms, but higher order of approximation is required in a theory to achieve a good quantitative agreement.

\end{abstract}

\date{\today}
\maketitle

\section{\label{I}Introduction}

Polymer macromolecules of complex branched structure are widely encountered in industry, biology and medicine \cite{Hadjichristidis2011}. Star polymers are the simplest representatives of this class of molecular architectures, consisting of a number $f$ (functionality) of linear branches radiating from a central core. They represent the model soft ``hybrid'' spheres encompassing both polymeric (small $f$) and colloidal (large $f$) type of behavior \cite{Vlassopoulos2001}. Advances in macromolecular chemistry have led to the anionic synthesis of nearly monodisperse model star homopolymers with high functionality $f$ (up to 128 branches) \cite{Toporowski1986}.

\begin{figure}[b!]
\begin{center}
\includegraphics[width=0.18\textwidth,angle=250]{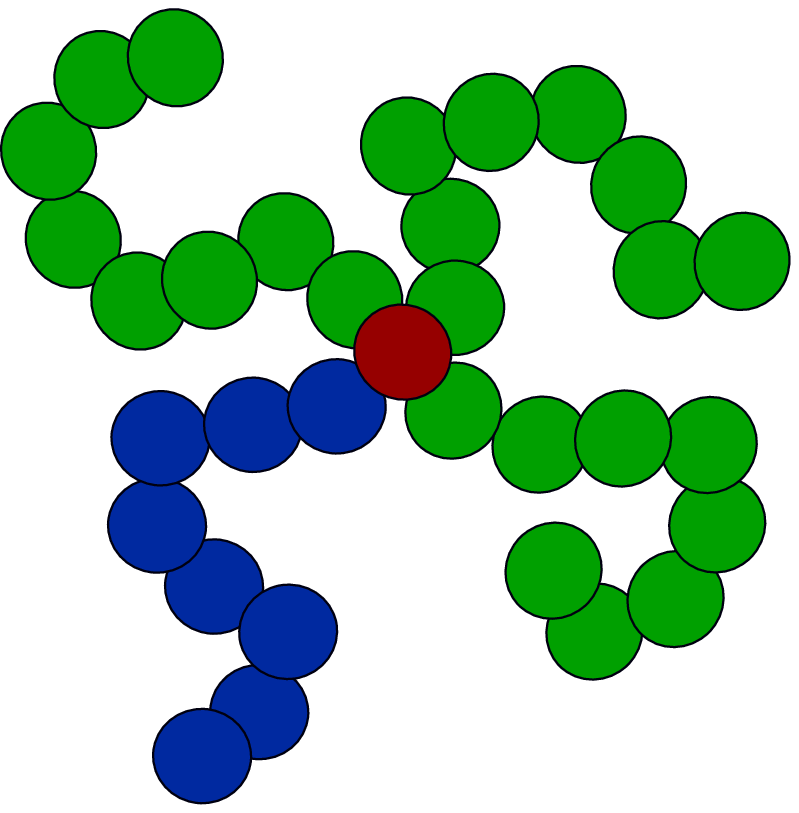}$\qquad\qquad\qquad$\includegraphics[width=0.3\textwidth,angle=270]{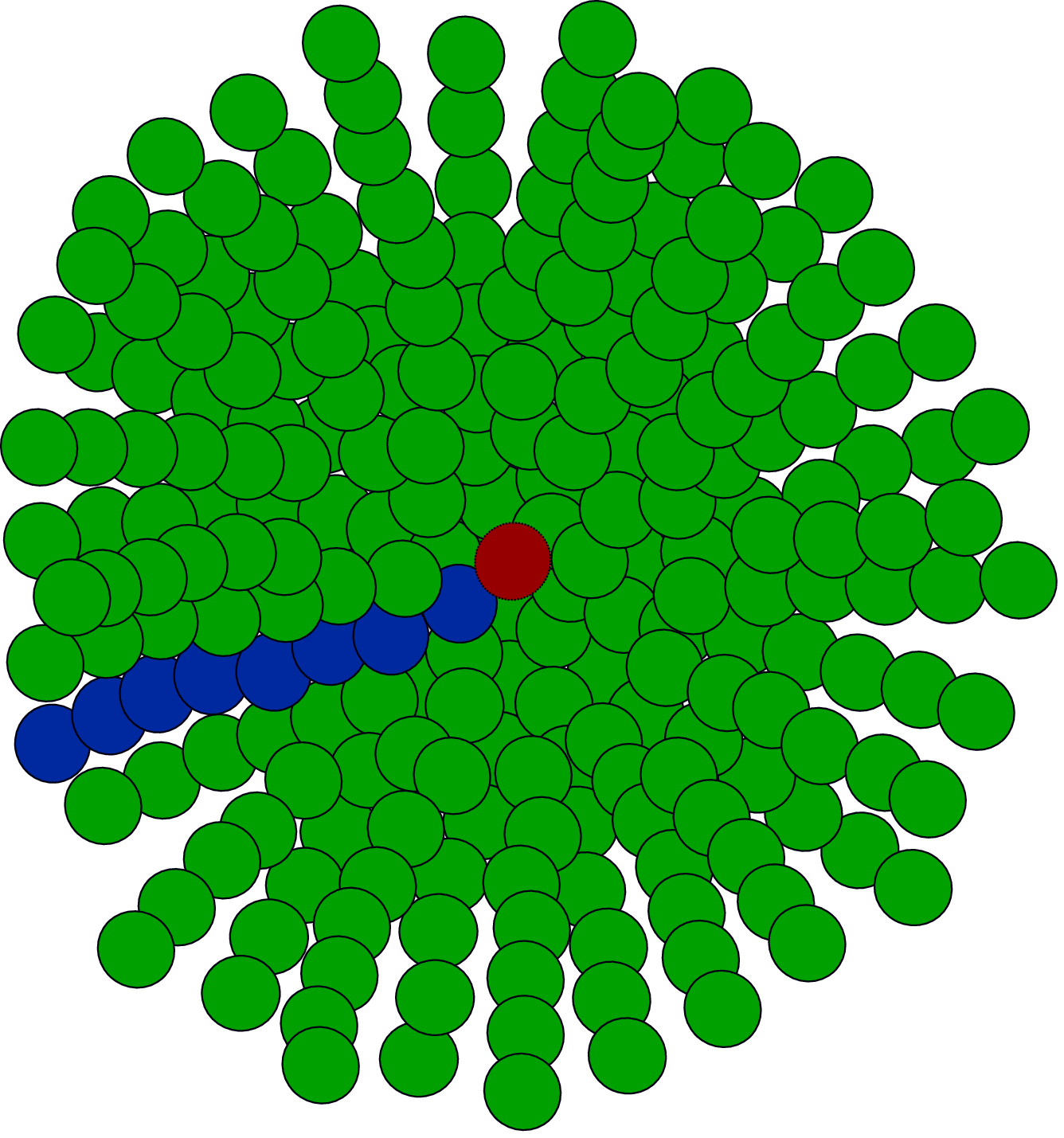}
\end{center}
\caption{\label{scheme} Schematic presentation of single arm conformation in branched $f$-arm polymer structure at polymeric (left) and colloidal (right) regimes. }
\end{figure}

Star-shaped polymers have lower viscosities compared to their linear counterparts of the same molecular weight in both solution and melt, and are widely encountered in bio-applications due to their good solubility and excellent bio-compatibility \cite{Zhou10,Tang16}. Amphiphilic star-shaped copolymers have recently attracted much attention because they can assemble into micelles with cores surrounded by hydrophilic shells in aqueous medium \cite{Heise99,Jin18,Meier05,Zhu06}. The encapsulation of guest molecules into these shells can be used, in particular, in controlled delivery of drugs \cite{Liu00,Haag04,Wang12,Soliman11,Chen15} or fluorescent sensors for metal cations \cite{Fernandez04}. The synthesis of star polymer nanoparticles is an active field of research \cite{Li12,McKenzie15}. On the other hand, star-like polymer structures are encountered in industry being the building blocks of polymer hydrogels \cite{Keys98,Dalton08} and can be in general considered as parts of polymer networks of more complicated topology \cite{Duplantier89,Schfer92}.

Star polymers are highly tunable in terms of the chemical composition of their arms, their number and length. Given the other factors are equal, the increase of the number of arms results in a gradual change of their properties from these resembling a group of weakly coupled linear chains to that of a compact colloid-like object. Such transformation, in turn, affects the solvability, rheological behavior and the ability to work as a host in the host-guest chemistry applications. On the level of the individual arms, such change manifests itself in the conformation changes from a coil-like to a more stretched state, recalling some features of dense brushes (see Fig. \ref{scheme}). Therefore, the study of these conformation changes may shed some more light on the internal structure and behavior of star polymers relevant to practical application.

To make the following discussion more specific, we will introduce a number of common characteristics for the polymer size and shape. Averaged (over available molecules and time trajectory) squared gyration radius of the star-polymer with $f$ branches each containing $N$ monomers is given by
\begin{equation}\label{R2gf}
\langle R^2_{g,f} \rangle =  \left\langle\frac{1}{1+Nf}\sum_{n=1}^{1+Nf}[\vec{r}_{n} - \vec{R}_c]^2\right\rangle,
\end{equation}
where $\vec{r}_{n}$ are coordinates of $n$th monomer and $\vec{R}_c$ is the center of mass of a star. The shape factor
\begin{equation}\label{gf}
g(f)=\frac{\langle R^2_{g,f}\rangle}{\langle R^2_{g,1}\rangle},
\end{equation}
first introduced by Zimm and Stockmeyer in 1949 \cite{Zimm1949}, characterizes the level of ``compactization'' of the star in comparison with an equivalent mass linear chain. Hereafter the index $f$ will denote the property of a star with $f$ arms, whereas the index $1$ -- the property of a respective linear object with the same number of monomers. Besides the effect of compactization, the star-polymer is also expected to become more spherical with the increase of $f$, as compared to the linear polymer, the associated property that characterizes the level of ``spherization'' of the star can be defined as \cite{Batoulis1989}
\begin{equation}\label{pAf}
p_A(f)=\frac{\langle A_f\rangle}{\langle A_1\rangle},
\end{equation}
where $A_f$ is the asphericity of a star with $f$ branches and $A_1$ -- the same for an equivalent mass linear chain.

Yet another characteristic, specific to a star-polymer, is the average center-end distance $\langle R^2_{ce,f}\rangle$ equivalent to the average square end-to-end distance of individual arms $r^2_{e,f}$:
\begin{equation}\label{R2cef}
\langle R^2_{ce,f} \rangle \equiv \langle r^2_{e,f} \rangle = \left\langle \frac{1}{f}\sum_{k=1}^{f}[\vec{r}_{k,N} - \vec{r}_c]^2\right\rangle.
\end{equation}
Here $\vec{r}_{k,i}$ ($i=1,\ldots,N$) is the position of $i$th
monomer within $k$th arm and $\vec{r}_c$ is the position of a
central monomer of a star. Some studies \cite{Grest1987} also
provide the average squared gyration radius of individual arms
\begin{equation}\label{arm_r2gf}
\langle r^2_{g,f} \rangle = \frac{1}{f}\sum_{k=1}^{f}\frac{1}{N}\sum_{i=1}^{N}[\vec{r}_{k,i} - \vec{r}_{k,c}]^2,
\end{equation}
where $\vec{r}_{k,c}$ is the center of mass of $k$th arm. Similarly to Eq.~(\ref{gf}), the dimensionless arm stretch ratio $p_e(f)$ and arm swelling ratio $p_g(f)$ can be introduced:
\begin{equation}
p_e(f) = \frac{\langle R^2_{ce,f}\rangle}{\langle r^2_{e,1}\rangle} \equiv \frac{\langle r^2_{e,f}\rangle}{\langle r^2_{e,1}\rangle},~~
p_g(f) = \frac{\langle r^2_{g,f}\rangle}{\langle r^2_{g,1}\rangle},\label{pef_pgf}
\end{equation}
where $r^2_{e,1}$ and $r^2_{g,1}$ are the square end-to-end distance and the square gyration radius, respectively, of a freely suspended linear chain with the length $N$ equal to that of a single arm within a star. There ratios characterize the averaged effect on a size of an arm due to the presence of adjacent branches. Finally, the arm asphericity ratio $p_a(f)$ between the asphericity of an individual arm within a star, $a_{f}$, and that of the same length chain $a_1$ freely suspended in a solvent, can be defined as
\begin{equation}\label{paaf}
p_a(f)=\frac{\langle a_f\rangle}{\langle a_1\rangle}.
\end{equation}

Let us stress that for the case of the Gaussian model, the monomers has zero size and the branches are non-interacting random walks. Therefore, for $g(f)$ one has the result obtained in Ref.~\cite{Zimm1949}, whereas the result for $p_e(f)$ and $p_g(f)$ is trivial:
\begin{equation}\label{gf_gauss}
g(f)=\frac{3f-2}{f^2},~~p_e(f)=1,~~p_g(f)=1. \label{ratios_Gauss}
\end{equation}
This means that for any $f \geq 3$, $g(f)<1$ and $g(f)\to 0$ at $f\to\infty$. This is an example of universal characteristics in statistics of polymers, which are independent on any details of chemical structure and are governed only by so-called global parameters, such as dimension of space $d$ or functionality $f$. Dimensional dependence of the size ratio $g(f)$ is found by introducing the concept of excluded volume, which refers to the idea that any segment (monomer) of macromolecule is not capable of occupying the space that is already occupied by another segment. Later on, analytical~\cite{Miyake84,Alessandrini92} and numerical~\cite{Batoulis89,Bishop93,Wei97} studies have found the value of $g(f)$ to increase if the excluded volume effect is taken into consideration. Concerning other characteristics of a non-Gaussian star, it has been shown by Duplantier \cite{Duplantier89}, that the scaling laws
\begin{equation}\label{Rce_scaling}
\langle R^2_{g,f} \rangle \sim \langle R^2_{ce,f} \rangle \sim K_f N^{2\nu},
\end{equation}
similar to that of a linear polymer of the length $N$, hold. Here the exponent $\nu$ is universal, but not the amplitude $K_f$ termed often as a ``swelling factor'' which incorporates the excluded volume effects. Its behavior is of much interest at both finite $f$ and at $f\to\infty$ \cite{Hsu2004}.

Some insight on the behavior of $K_f$ as function of $f$ is provided by the scaling ansatz of Daoud and Cotton \cite{Daoud82,Birsh84} obtained within a framework of a ``blob model''. A star there is represented by an inner meltlike ``extended core region'', an intermediate region resembling a concentrated solution, and an outer semidilute region. As was shown in Ref.~\cite{Grest1987}, at $d=3$ this ansatz leads to the scaling of a form $K_f \sim f^{1-\nu}$, and substituting this form into Eqs.~(\ref{gf}) and (\ref{pef_pgf}) yields
\begin{equation}\label{g_pe_Daoud}
g(f) \sim f^{1-3\nu},~~ p_e(f) \sim f^{1-\nu}.
\end{equation}
Available  results of Monte Carlo \cite{Batoulis89,Zifferer1999,Havrnkova03,Hsu2004} and molecular dynamics \cite{Grest1987,Grest1994} simulations, provide a range of different values for the exponents in Eq.~(\ref{g_pe_Daoud}), which tend to move toward the Daoud and Cotton scaling law at larger $f$ and $N$. Having enough statistics in this regime, however, is difficult due to long relaxation times. The shape characteristics such as asphericity and prolateness received some attention as well \cite{Batoulis89,Zifferer1999} but to a lesser extent.

As far as there are discrepancies in available values for the universal ratios of the size and a form of star-polymers, application of different theoretical and simulation techniques should be always welcome as these may shed some new information there. The aim of this study is two-fold. First, we perform theoretical study  based on Edwards continuous chain model description of star-like macromolecule. The universal size and shape characteristics of interest
are estimated  applying the direct polymer renormalization scheme developed in the works of des Cloizeaux \cite{desCloiseaux}. This method has been proved to be efficient in analyzing the size and shape properties of molecules of various architecture in the asymptotic limit
of infinitely long chains (see e.g. \cite{Blavatska12,Haydukivska14,Blavatska15}). Then, the mesoscopic dissipative particle dynamics (DPD) based simulations are undertaken. Earlier such simulation have been employed to analysis of the size \cite{Ilnytskyi2007} and shape \cite{Kalyuzhnyi2016} characteristics of a linear chain, as well as the shape properties of heterostar polymers \cite{Kalyuzhnyi2018}. The main point of interest is whether or not such simulations, that involve soft interaction potential, can reproduce the effect of branches stretch and swelling upon the increase of their number $f$. To this end we compare our findings with available simulation data performed on the lattice models of polymers and the van der Waals potential based molecular dynamics studies. For the shape characteristics missing in the previous Monte Carlo simulations studies, we undertake our own simulations using this method.

The outline of the study as follows. Section \ref{II} contains theoretical results based  on direct polymer renormalization, section \ref{III} explains the simulation approaches being used in this study and presents the results of the computer simulations, conclusions are given in section \ref{IV}.

\section{\label{II}Theory}
\subsection{Continuous Chain Model}

In the frames of continuous chain model \cite{Edwards}, each arm of
the star is presented by a path $ r_i(s)$, parameterized by $0\leq
s\leq L_i$, $i=1,2,\ldots,f$.  We take the contour length of all
arms to be equal: $L_1=\ldots =L_f=L$. The hamiltonian of the system
reads:
\begin{eqnarray}
H = \frac{1}{2}\sum_{i=1}^{f}\,\int_0^L ds\,\left(\frac{d\vec{r_i}(s)}{ds}\right)^2+\frac{u}{2}\sum_{i,j=1}^{f}\int_0^Lds'\int_0^L ds''\,\delta(\vec{r_i}(s')-\vec{r_j}(s'')),\label{H}
\end{eqnarray}
where the first term  describes the chain connectivity and the second one corresponds to excluded volume interaction with coupling constant $u$.
Note that performing the dimensional analysis of couplings in above expression yields $[u]=[L]^{(4-d)/2}$. The ``upper critical'' value
of space dimension $d_c=4$, at which the coupling becomes dimensionless, is important in developing the renormalization scheme, as will be shown in next section.

 The star-like architecture can be described as a set of chains, connected by their end points, so that the partition function of such a system reads:
 \begin{eqnarray}
Z^{{\rm star}}_{f}=\frac{1}{Z_0^{{\rm star}}}\int\,D\vec{r}(s)\,\prod_{i=1}^{f}\delta(\vec{r_i}(0))\,{\rm e}^{-H}.
\label{Zs}
\end{eqnarray}
Here, a multiple path integral is performed for the paths $r_1,\ldots,r_f$,
the product of $\delta$-functions reflects the star-like configuration of $f$ chains, and $Z_0$ is a partition function of a Gaussian molecule:
 \begin{equation}
 Z_0^{{\rm star}}=\int\,D\vec{r}(s)\,\prod_{i=1}^{f}\delta(\vec{r_i}(0)){\rm e}^{-\frac{1}{2}\sum_{i=1}^{f}\,\int_0^L ds\,\left(\frac{d\vec{r_i}(s)}{ds}\right)^2}.
 \end{equation}

\subsection{Direct Renormalization}

On the base of continuous chain model the observables are expressed as functions that depend on the individual
chain length $L$ and diverge in the limit $L\rightarrow \infty$. These divergences need to be eliminated
 to obtain the universal values of the parameters under consideration. For that  purpose the direct renormalization method was developed by des Cloiseaux \cite{desCloiseaux}. The key point of the method is to introduce the
  renormalization factors that are directly connected with  the physical quantities and
   allow to remove the divergences. In order to obtain the finite values of observables one needs to evaluate them at the corresponding
   fixed points (FPs) of the renormalization group.

 The task of calculating the FP starts from obtaining the partition function of two interacting polymers $Z(L,L)$. It is also important to introduce the renormalization factors that are $[Z(L,u_0)]^{-2}$, with $Z$ being the partition function of a single chain and $\chi_0(L,\{x_0\})$ --  a so-called swelling factor  connected to end-to-end distance $\chi_0=\langle R_e^2\rangle/L$. A renormalized coupling constant can be presented as:
\begin{eqnarray}
u_R(u_0)= - [Z(L,u_0)]^{-2}Z(L,L)[2\pi \chi_0(L,u_0)]^{-2+\epsilon/2}
\end{eqnarray}
where $\epsilon=4-d$ is a deviation from upper critical dimension for the coupling constant $u_0$. Taking a limit of infinitely long chain in this expression one can receive the fixed values of coupling constants:
\begin{eqnarray}
\lim_{L\to\infty} u_{R}(u_0)=u_R^*.
\end{eqnarray}
An interesting and well known fact about FPs is that their values do not depend on the topology of the macromolecule, and as a result can be calculated in the simplest case of single linear chain. Thus, we have \cite{desCloiseaux}:
\begin{eqnarray}
u^*_{{\rm Gauss}}=0,\,\,\,\,{\rm at\,}d\geq 4,\label{FPG}\\
u=\frac{\epsilon}{8},\,\,\,\,{\rm at\,}d< 4.\label{FPEV}
\end{eqnarray}
Here, (\ref{FPG}) corresponds to the polymer chain without interactions between monomers, whereas (\ref{FPEV}) describes the excluded volume effect.

\subsection{Results and Discussion}

Following the general scheme of \cite{desCloiseaux}, we evaluate
 all the observables of interest as perturbation theory series in an excluded volume coupling constant $u$ and
  restrict ourselves to the first order of expansion in $u$.
An analytical expression for partition function in this approximation  reads \cite{Blavatska}:
\begin{eqnarray}
 Z^{star}_f(L)=Z^{star}_0(L) \left\{1-\frac{4u}{(2-d)(4-d)} \right.\left. \left[f+ \frac{f(f-1)}{2}(2^{2-d/2}-2)\right]\right\},
\label{Zf}
\end{eqnarray}
where dimensionless coupling constant is introduced:
$u=u(2\pi)^{-d/2}L^{2-d/2}$.

Finally, one may perform a $\epsilon$-expansions:
\begin{eqnarray}
 Z^{star}_f=1+u\left[\frac{f(3-f)}{\epsilon}+\frac{f(3-f)}{2}+\right.
\left.\frac{f(f-1)}{2}\ln 2\right].
\label{Zfeps}
\end{eqnarray}
In what follows, the partition function (\ref{Zfeps}) will be used in evaluation the averaged values of observables under interest,
defined by:
\begin{eqnarray}
\langle (\ldots) \rangle = \frac{1}{{ Z^{star}_{f}(L)}}\prod_{i=1}^{f}\int\,D\vec{r}(s)
\delta(\vec{r_i}(0)){\rm e}^{-H}(\ldots).
\end{eqnarray}

{\bf{2.3.1 Center-end Distance of the Individual Arm in Star Structure}}

We start with considering the center-end distance of randomly chosen arm in a star defined by (\ref{R2cef}) also in the case of continuous chain model it is more practical to present it as:
\begin{eqnarray}
\langle r^2_{{ e,f}}\rangle = \frac{1}{f}\sum_{i=1}^{f}{\langle(\vec{r}_i(L)-\vec{r}_i(0))^2\rangle}.
\end{eqnarray}
We obtain:
\begin{eqnarray}
\langle r_{{ e, f}}^2\rangle=Ld\left(1+u\left[\left(\frac{2}{\epsilon}-1\right)+(f-1)\left(\ln 2 -\frac{1}{4}\right) \right]\right).
\end{eqnarray}

Recalling that the end-to-end distance of an individual chain reads:
\begin{eqnarray}
\langle r_{{ e, 1}}^2\rangle=Ld\left(1+u\left(\frac{2}{\epsilon}-1\right)\right),
\end{eqnarray}
we obtain for the stretch ratio $p_e(f)$ given by Eq. (\ref{pef_pgf}):
\begin{equation}
p_{{\rm e}}(f)=\frac{\langle r_{{ e, f}}^2\rangle}{\langle r_{{ e, 1}}^2\rangle}=1+ u(f-1)\left(\ln 2 -\frac{1}{4}\right).
\end{equation}
Substituting the fixed points values (\ref{FPG}), (\ref{FPEV}) into this expression, we obtain:
\begin{eqnarray}
&&p_{{\rm e}}^{{\rm Gauss}}=1,\\
&&p_{{\rm e}}(f)=1+ \frac{\epsilon}{8}(f-1)\left(\ln 2 -\frac{1}{4}\right). \label{eEV}
\end{eqnarray}

{\bf{2.3.2 Gyration Radius of the Individual Arm in Star Structure}}

Within the continuous chain model the gyration radius of an individual arm can be presented as:
\begin{eqnarray}
\langle r^2_{{g, f}}\rangle = \frac{1}{f}\sum_{i=0}^{f}\frac{1}{2L^2}
\int_0^L\int_0^L ds_1\,ds_2 \langle(\vec{r}_i(s_2)-\vec{r}_j(s_1))^2\rangle.
\end{eqnarray}
We obtain:
 \begin{eqnarray}
\langle r^2_{{ g, f}}\rangle = \frac{dL}{6}\left( 1+u\left[ \left( \frac{2}{\epsilon}-\frac{13}{12}\right)\right.\right.\left.\left.+(f-1)\left(\frac{35}{8}-6\ln(2)\right) \right]\right).
\end{eqnarray}
Remembering that the radius of gyration of an individual chain reads:
\begin{eqnarray}
\langle r_{{ g, 1}}^2\rangle=\frac{Ld}{6}\left(1+u\left(\frac{2}{\epsilon}-\frac{13}{12}\right)\right),
\end{eqnarray}
we obtain for the arm swelling ratio $p_g(f)$ given by Eq. (\ref{pef_pgf}):
\begin{equation}
p_{{\rm g}}(f)=\frac{\langle r_{{ g, f}}^2\rangle}{\langle r_{{ g, 1}}^2\rangle}=1+ u(f-1)\left(\frac{35}{8}-6\ln 2 \right) .
\end{equation}
Substituting the fixed points values (\ref{FPG}), (\ref{FPEV}) into this expression, we obtain:
\begin{eqnarray}
&&p_{{\rm g}}^{{\rm Gauss}}=1,\\
&&p_{{\rm g}}(f)=1+ \frac{\epsilon}{8}(f-1)\left( \frac{35}{8}- 6\ln 2 \right) . \label{gEV}
\end{eqnarray}

{ \bf{2.3.3 Asphericity of the Individual Arm of Star}}

In the terms of model under consideration the gyration tensor of an arm in a star can be presented as:
\begin{eqnarray}
Q_{\alpha,\beta} = \frac{1}{f}\sum_{i=0}^{f}\frac{1}{2L^2}
\int_0^L\int_0^L ds_1\,ds_2 (\vec{r}^{\alpha}_i(s_2)-\vec{r}^{\alpha}_j(s_1))
(\vec{r}^{\beta}_i(s_2)-\vec{r}^{\beta}_j(s_1)).
\end{eqnarray}

To characterize the deviation of a shape from the sphere it is useful to calculate the asphericity that in terms of the components of gyration tensor can be presented as \cite{Aronovitz}:
\begin{eqnarray}
{\hat {a}}_f = \frac{\langle Q_{\alpha,\alpha}^2\rangle+d\langle Q_{\alpha,\beta}^2\rangle-\langle Q_{\alpha,\alpha}Q_{\beta,\beta}\rangle}{\langle Q_{\alpha,\alpha}^2\rangle+(d-1)\langle Q_{\alpha,\alpha}Q_{\beta,\beta}\rangle}.
\end{eqnarray}
In one loop approximation we obtain:
\begin{eqnarray}
\hat{a}(f) =\frac{1}{2}+\frac{1}{48}\,\epsilon+u \left( \frac {805}{8}\,(f-1)\ln  2 \right.\left.  -{\frac {53509}{768}}\,f+\frac {53561}{768} \right).
\end{eqnarray}

Recalling that the asphericity of an individual chain reads:
\begin{eqnarray}
&&\hat{a}_1 =\frac{1}{2}+\frac{1}{48}\,\epsilon+u \frac {13}{192},
\end{eqnarray}
we obtain for the asphericity ratio $p_a(f)$  given by Eq. (\ref{paaf}):
\begin{eqnarray}
&&p_a(f)=\frac{\hat{a}_{{ f}}}{\hat{a}_1}=1+u(f-1)\left(\frac {805}{4}\ln 2 \right.\left.  -{\frac {53509}{384}} \right).
\end{eqnarray}

In fixed points (\ref{FPG}), (\ref{FPEV}) we obtain
\begin{eqnarray}
&&p_a^{Gauss} =1,\\
&&p_a(f) =\frac{1}{2}+\frac{\epsilon}{8}(f-1)
\left(\frac {805}{4}\,\ln  2 \right.\left.  -{\frac {53509}{384}}\right). \label{aEV}
\end{eqnarray}

\section{\label{III}Computer simulations}

\subsection{\label{III_1}Details of the dissipative particle dynamics simulations}

The mesoscopic simulations technique have been used to test the theory. In our study we follow DPD method as formulated in Ref.~\cite{Groot1997}. In this approach the polymer and solvent molecules are modeled as soft beads of equal size, each of these beads represents a group of atoms. The following reduced quantities are used: the length will be measured in units of the diameter of a soft bead, and the energy in units of $k_{B}T$. Here, $k_{B}$ is the Boltzmann constant and $T$ is the temperature. In a polymer chain monomers are connected via harmonic springs, which results in a force:
\begin{equation}\label{FB}
  \vvec{F}^B_{ij} = -k\vvec{x}_{ij}\,,
\end{equation}
where $k$ is the spring constant, and $\vvec{x}_{ij}=\vvec{x}_i-\vvec{x}_j$, $\vvec{x}_i$ and $\vvec{x}_j$ are the coordinates of $i$th and $j$th bead, respectively. The non-bonding force $\vvec{F}_{ij}$ acting on the $i$th bead from his $j$th counterpart is expressed as a sum of three contributions
\begin{equation}
  \vvec{F}_{ij} = \vvec{F}^{\mathrm{C}}_{ij} + \vvec{F}^{\mathrm{D}}_{ij}
  + \vvec{F}^{\mathrm{R}}_{ij}\,,
\end{equation}
where $\vvec{F}^{\mathrm{C}}_{ij}$ is the conservative force that defines the repulsion between the beads, $\vvec{F}^{\mathrm{D}}_{ij}$ is the dissipative force which is responsible for the friction between them and the random force $\vvec{F}^{\mathrm{R}}_{ij}$  works in pair with a dissipative force to thermostat the system. The expression for all these three contribution are given below \cite{Groot1997}
\begin{equation}\label{FC}
  \vvec{F}^{\mathrm{C}}_{ij} =
     \left\{
     \begin{array}{ll}
        a(1-x_{ij})\displaystyle\frac{\vvec{x}_{ij}}{x_{ij}}, & x_{ij}<1,\\
        0,                       & x_{ij}\geq 1,
     \end{array}
     \right.
\end{equation}
\begin{equation}\label{FD}
  \vvec{F}^{\mathrm{D}}_{ij} = -\gamma
  w^{\mathrm{D}}(x_{ij})(\vvec{x}_{ij}\cdot\vvec{v}_{ij})\frac{\vvec{x}_{ij}}{x^2_{ij}},
\end{equation}
\begin{equation}\label{FR}
  \vvec{F}^{\mathrm{R}}_{ij} = \sigma
  w^{\mathrm{R}}(x_{ij})\theta_{ij}\Delta t^{-1/2}\frac{\vvec{x}_{ij}}{x_{ij}},
\end{equation}
where $a$ is the amplitude for the conservative repulsive force, $x_{ij}=|\vvec{x}_{ij}|$, $\vvec{v_{ij}}=\vvec{v}_{i}-\vvec{v}_{j}$, $\vvec{v}_{i}$ is the velocity of the $i$th bead. The dissipative force has an amplitude $\gamma$ and decays with distance according to the weight function $w^{\mathrm{D}}(x_{ij})$. The amplitude for the random force is $\sigma$ and the respective weight function is $w^{\mathrm{R}}(x_{ij})$. $\theta_{ij}$ is the Gaussian random variable. As was shown by Espa\~{n}ol and Warren \cite{Espanol1995}, to satisfy the detailed balance requirement, the amplitudes and weight functions for the dissipative and random forces should be interrelated: $\sigma^2=2\gamma$ and $w^{\mathrm{D}}(x_{ij})=\left[w^{\mathrm{R}}(x_{ij})\right]^2$. Here we use quadratically decaying form for the weight functions:
\begin{equation}
w^{\mathrm{D}}(x_{ij})=\left[w^{\mathrm{R}}(x_{ij})\right]^2
=\left\{
\begin{array}{ll}
(1-x_{ij})^2, & x_{ij} < 1,\\
0, & x_{ij} \geq 1.
\end{array}
\right.
\end{equation}
The reduced density of the system is defined as $\rho^{*} = (fN + N_{s})/V=3$ , where $N_s$ is the number of solvent particles and $V$ is system volume. The other parameters are chosen as follows: $\gamma=6.75$, $\sigma=\sqrt{2\gamma}=3.67$.

The focus of this study is the excluded volume related ``swelling'' of star polymer branches that takes place upon the increase of their number $f$. Scaling with respect to the arm length $N$ is not considered and we set $N=8$ in all cases. The number of branches $f$ ranged from $f=1$ (linear chain) to the maximum of $f=25$. The effect of solvent quality \cite{Grest1994,Kalyuzhnyi2018} is also left beyond our analysis and we concentrated on the case of ``athermal'' solvent only where $a_{pp}=a_{ss}=a_{ps}$.

One could argue that $N=8$ is quite a moderate length of individual branches comparing to some other simulation studies \cite{Grest1987,Grest1994,Zifferer1999,Hsu2004,Havrnkova03}, but let us not forget that the DPD approach considers polymer on a mesoscopic level. Therefore, one of the points of interest is its efficiency in reducing the required degrees of freedom to describe scaling properties of interest. For instance, scaling properties of a linear polymer chain can be reproduced fairly well using chain lengths of $10\leq N \leq 40$ \cite{Ilnytskyi2007,Kalyuzhnyi2016}.

Simulation length in the current study is $4 \cdot 10^6$ DPD steps for each case with the time-step of $0.04$ in reduced units, where the first $4\cdot 10^5$ steps are left for the equilibration. The rest of the time trajectory is used to evaluate the average values for the size properties $R^2_{g,f}$, $R^2_{ce,f}\equiv r^2_{e,f}$, $r^2_{g,f}$, as well as for the asphericities $A_f$ and $a_f$, and, finally, for their universal combinations $g(f)$, $p_A(f)$, $p_e(f)$, $p_g(f)$ and $p_a(f)$, all being defined in Sec.~\ref{I}.

\subsection{\label{III_1a}Details of the Monte Carlo simulations of a lattice model}

Although the main focus in the present paper is made on DPD simulations which is
proved to be more effective in dealing with branched structures with strong density fluctuations,
the part of our studies are performed for comparison also applying the lattice simulations.
We apply the algorithm of growing chain, which is based on ideas from the very
first days of Monte Carlo simulations, the Rosenbluth-Rosenbluth (RR) method \cite{Rosenbluth}.
Here,  the $n$th monomer of a chain is placed at one of $2d$ randomly chosen neighbor site
of the last placed $(n - 1)$th monomer ($n\leq N$, where N
is the total desired length of the chain, $d$ is dimension of the lattice).
If the chosen  site is already occupied by growing trajectory, then it is avoided. The weight $ W_n = \prod_{l=2}^n m_l
$ is given to each chain configuration at the $n$th step, here $m_l$
is the number of allowed (not-occupied)
lattice sites to place the $l$th monomer. When the chain
of a total length $N$ is constructed, the new trajectory starts to grow from
the same starting point, until the desired number of chain
configurations are obtained. The configurational averaging for any observable of interest $O$ is then given
by:
\begin{equation}
\langle O \rangle = \frac{\sum_{\rm conf} W_N^{{\rm conf}} O_{{\rm conf}}}{\sum_{\rm conf} W_N^{{\rm conf}}},
\end{equation}
where $W_N^{\rm conf}$
 is the weight of an $N$-monomer chain in a
given configuration.

This method can be easily generalized to the case of $f$-arm branched structure. In the present work, we introduced the central core in the form of
cube of the minimum size $3 \times 3 \times 3$, and let the sites on cube edges be the starting points of a set of growing chains trajectories.
$f$ chains are growing simultaneously, i.e. at each round the $n$th monomer of $f$th chain is added to $(n - 1)$ previously placed monomers of this chain,
taking into account, that all $f$ chains are avoiding each other.
In our study, we consider from $f=2$ to $f=14$ arms, the maximum  length of each arm $N=60$ and perform averaging over  $10^5$ configurations.

\subsection{\label{III_2}Comparison of the Dissipative Particle Dynamics Results with Other Simulation Approaches and Analytic Ansatzes}

Simulation results obtained in the current study are compared closely with a number of the previous results obtained by means of known ansatzes and by various simulation methods. To simplify referring to these, we introduce the following shortcuts to be used in the figures: (a): Monte Carlo \cite{Zifferer1999}, (b): Gaussian model, (c): Daoud-Cotton ansatz \cite{Daoud82}, (d): Monte Carlo \cite{Batoulis89}, (e): Monte Carlo \cite{Hsu2004}, (f): Molecular dynamics \cite{Grest1994}, (g): Monte Carlo \cite{Whittington1986}.

\begin{figure}[t!]
\begin{center}
\includegraphics[width=0.5\textwidth,angle=270]{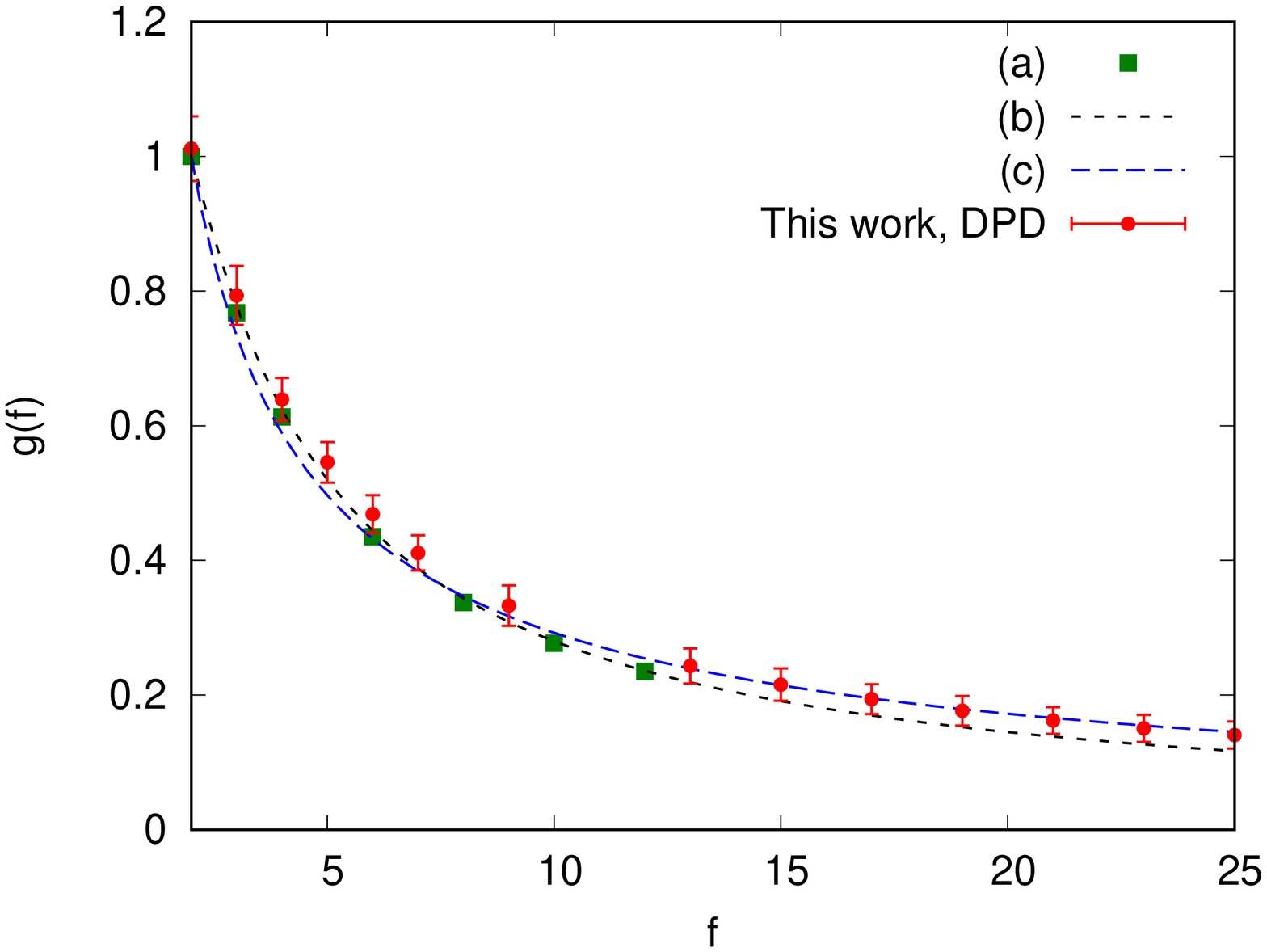}\\
\includegraphics[width=0.5\textwidth,angle=270]{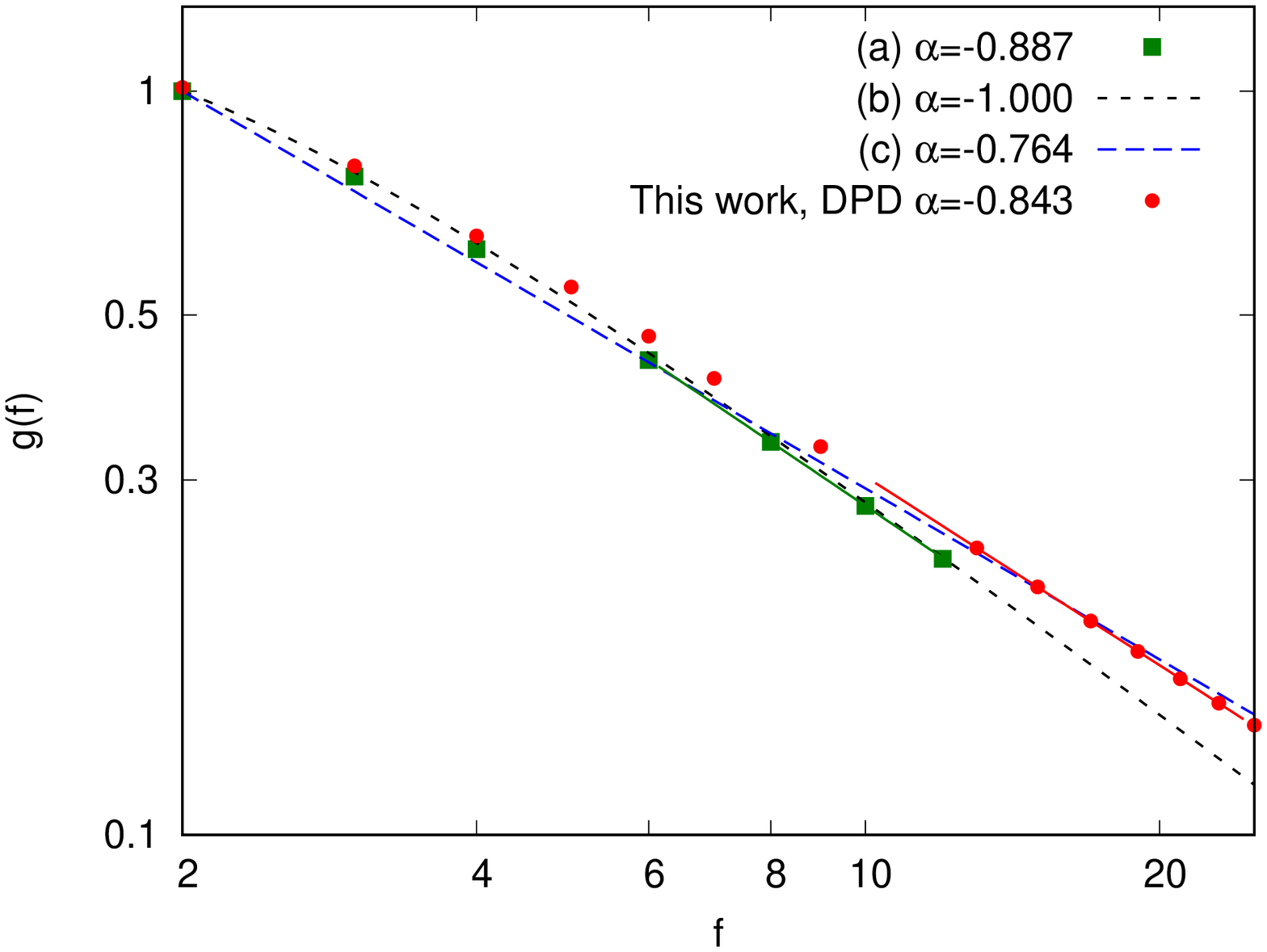}
\caption{\label{SAW_g-star} Top frame: the shape factor $g(f)$  as a function of the number of branches $f$, datapoints and curves are marked according to the shorthands given at the beginning of Sec.\ref{III_2}. Bottom frame contains the same data plotted in log-log scale yielding the estimates for the exponent $\alpha$ according to Eq.~(\ref{alpha}), fitting range is shown via straight lines.}
\end{center}
\end{figure}
We discuss the results for the shape factor $g(f)$ first. These are visualized in the top row of Fig.~\ref{SAW_g-star} where the DPD simulation results of this study (where $N=8$) are compared with the lattice Monte Carlo simulation study of Zifferer \cite{Zifferer1999} obtained by extrapolation to the case $N\to\infty$, as well as with the analytic results for the Gaussian model and in the frames of the Daoud-Cotton ansatz (\ref{g_pe_Daoud}). We were assuming a power law dependence for all data at $f \gg 1$
\begin{equation}\label{alpha}
g(f) \sim C f^\alpha,~~~f \gg 1,~~N=\mathrm{const}.
\end{equation}
an inherent feature of the Daoud-Cotton ansatz (\ref{g_pe_Daoud}), in which case $\alpha=1-3\nu\approx-0.764$ and $C=2^{-\alpha}$ (chosen to ensure that $g(2)=1$ for the linear chain case). To perform fitting of the data to the form (\ref{alpha}), we replot all of them in a logarithmic scale, as shown in the bottom row of Fig.~\ref{SAW_g-star}. For the Gaussian model one has $g(f)=(3f-2)/f^2$ \cite{Zimm1949}, hence $\alpha=-1$. Fitting the data of this study and that from Ref.~\cite{Zifferer1999} yields very close values of $\alpha=-0.843$ and $-0.887$, respectively, where the fitting ranges are provided by a solid color lines in Fig.~\ref{SAW_g-star}. The data comparison shown in a top row of Fig.~\ref{SAW_g-star} indicates very good agreement between the DPD simulations and the Daoud-Cotton ansatz at $10 \leq f \leq 25$, whereas Gaussian model demonstrates lower values of $g(f)$ in this interval of $f$. The simulation data of Ref.~\cite{Zifferer1999} is available only in the $3\leq f \leq 12$ range, where they are found to be closer to the result of the Gaussian model. One may remark that the DPD simulations, performed at a moderate arm length $N=8$, provide the shape factor $g(f)$ as a function of a number of branches $f$ which resembles very closely the results of the lattice Monte Carlo simulations \cite{Zifferer1999} at the $N\to\infty$ limit. This correlates well with the findings that a linear chain within the DPD enters scaling regime at moderate values of $N\sim 8-10$ \cite{Kalyuzhnyi2016}. It is also interesting to note that the Daoud-Cotton ansatz provide the result that is very close to our simulations at $f>10$, following the discussion in Ref.~\cite{Havrnkova03} that this ansatz is valid for the case of rather large number of branches $f$.

\begin{figure}[t!]
\begin{center}
\includegraphics[width=0.5\textwidth,angle=270]{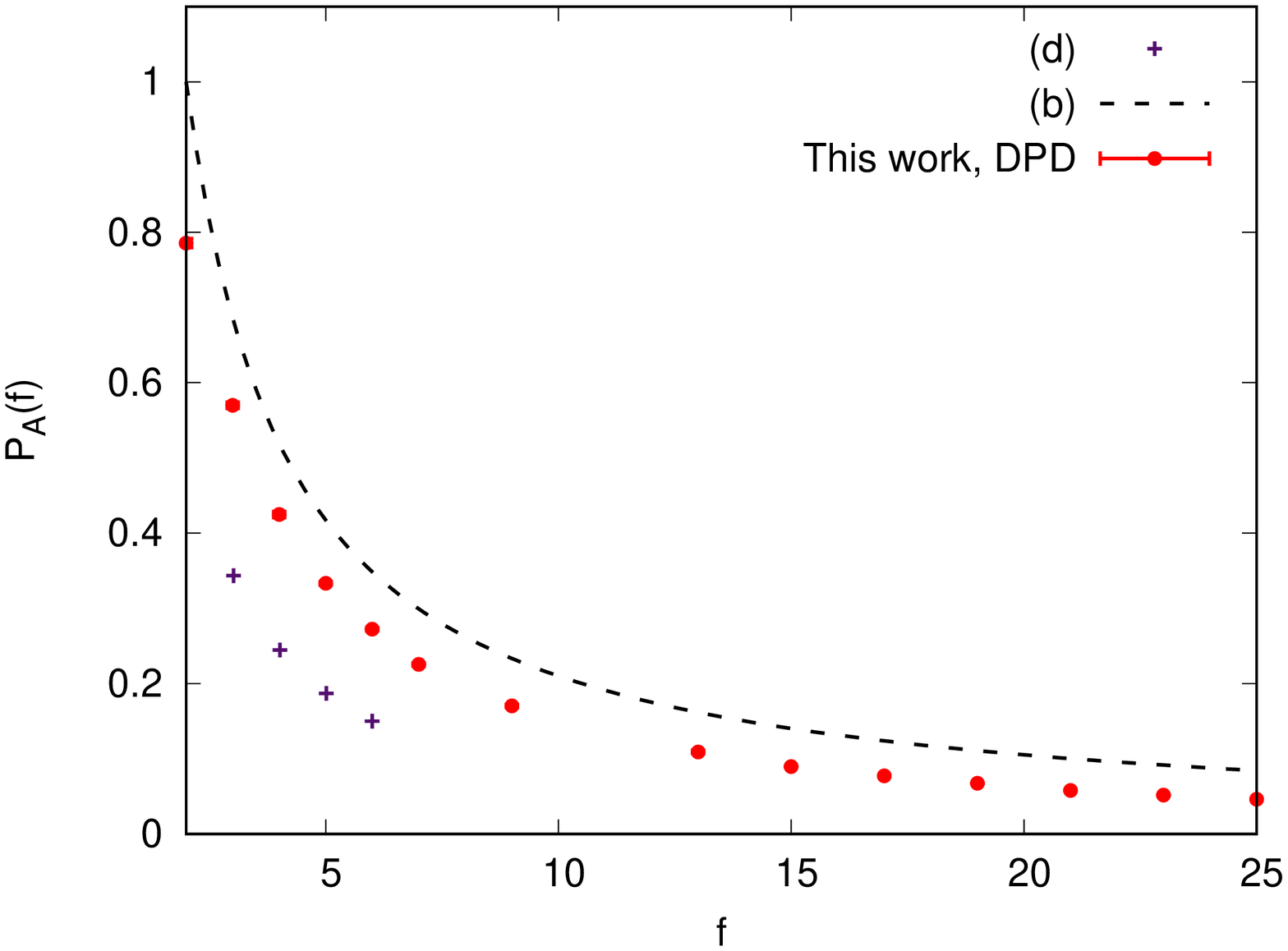}\\
\includegraphics[width=0.5\textwidth,angle=270]{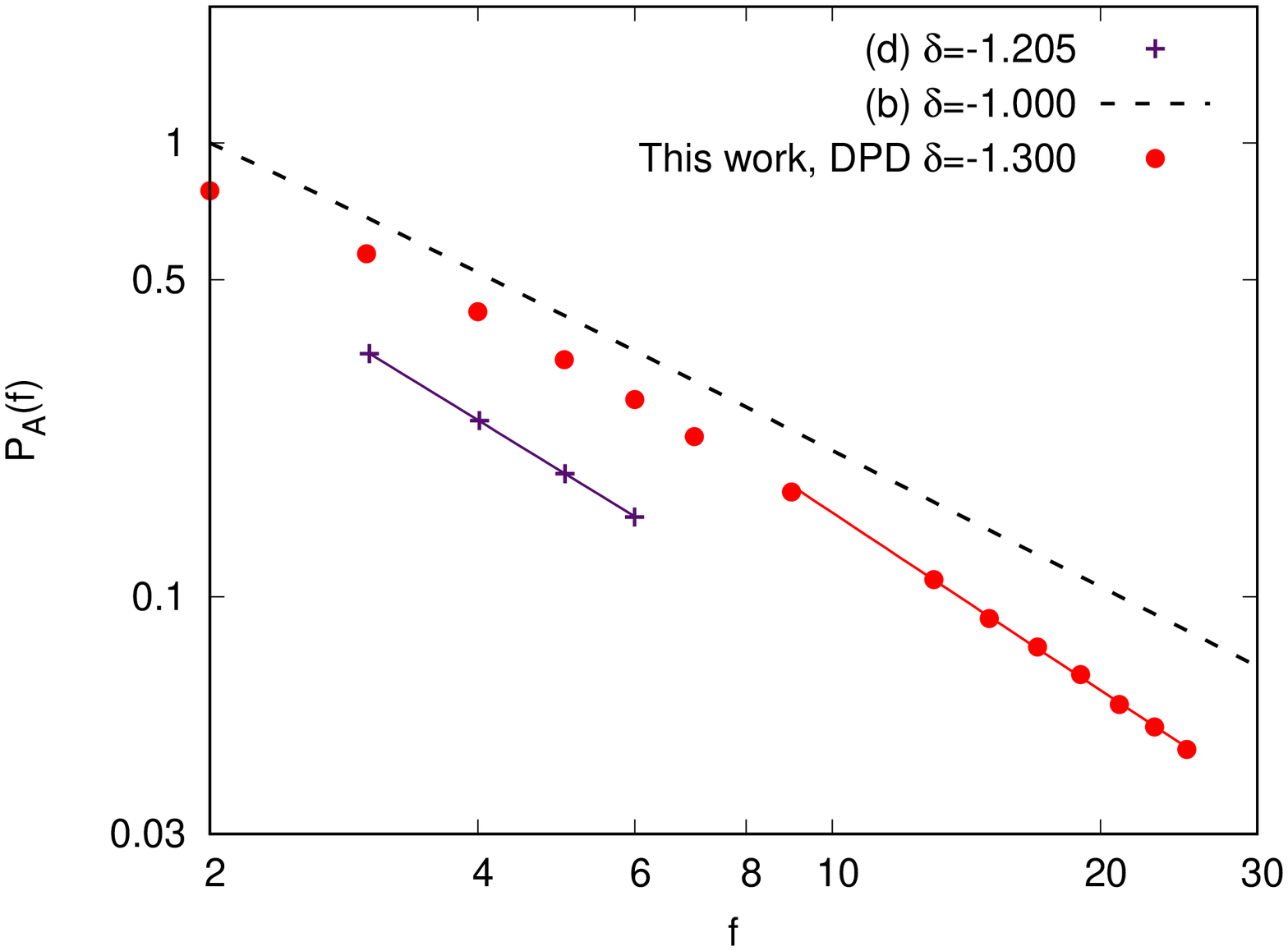}
\caption{\label{SAW_pA-star} The same as in Fig.~\ref{SAW_g-star} but for the star asphericity factor $p_A(f)$ (\ref{pAf}) and for the exponent $\delta$ in Eq.(\ref{delta}).}
\end{center}
\end{figure}

Let us now switch to the $p_A(f)$ ratio for a star defined in Eq.~(\ref{pAf}), the results obtained in this study are shown in Fig.~\ref{SAW_pA-star}, where they are compared to the simulations by Batoulis and Kremer \cite{Batoulis1989} and the results for the Gaussian model. The expression for the latter
\begin{equation}\label{pAfGauss}
\frac{57}{5}\frac{15f-14}{15(3f-2)^2+4(15f-14)}\sim\frac{1}{f},~~f\to\infty
\end{equation}
for the three-dimensional case is obtained from the relations provided in Ref.~\cite{Blavatska2015}. It is seen from Fig.~\ref{SAW_pA-star} that the excluded volume effects, present in both simulation studies, promote faster decay of $p_A(f)$ with the increase of $f$ than predicted by a Gaussian model. In simple words,
excluded volume star polymer became more spherical compared to the equivalent linear chain at smaller $f$ than their random walk counterparts. Logarithmic scale plots in Fig.~\ref{SAW_pA-star} indicate that the slope for $p_A(f)$ is getting linear with the increase of $f$, therefore, we may assume a power law dependence at large $f$
\begin{equation}\label{delta}
p_A(f) \sim D f^\delta,~~~f \gg 1,~~N=\mathrm{const}.
\end{equation}
The value of $\delta=-1.314$ is reported in this study and is compared against $-1.205$ obtained from fitting the data from Ref.~\cite{Batoulis1989}, however, the latter are restricted to the maximum value of $f=6$ and may underestimate the magnitude of this exponent at large $f$. In any case, the exponent $\delta$ in both simulations is found to deviate essentially from its Gaussian value $1$ (\ref{pAfGauss}).

\begin{figure}[t!]
\begin{center}
\includegraphics[width=0.5\textwidth,angle=270]{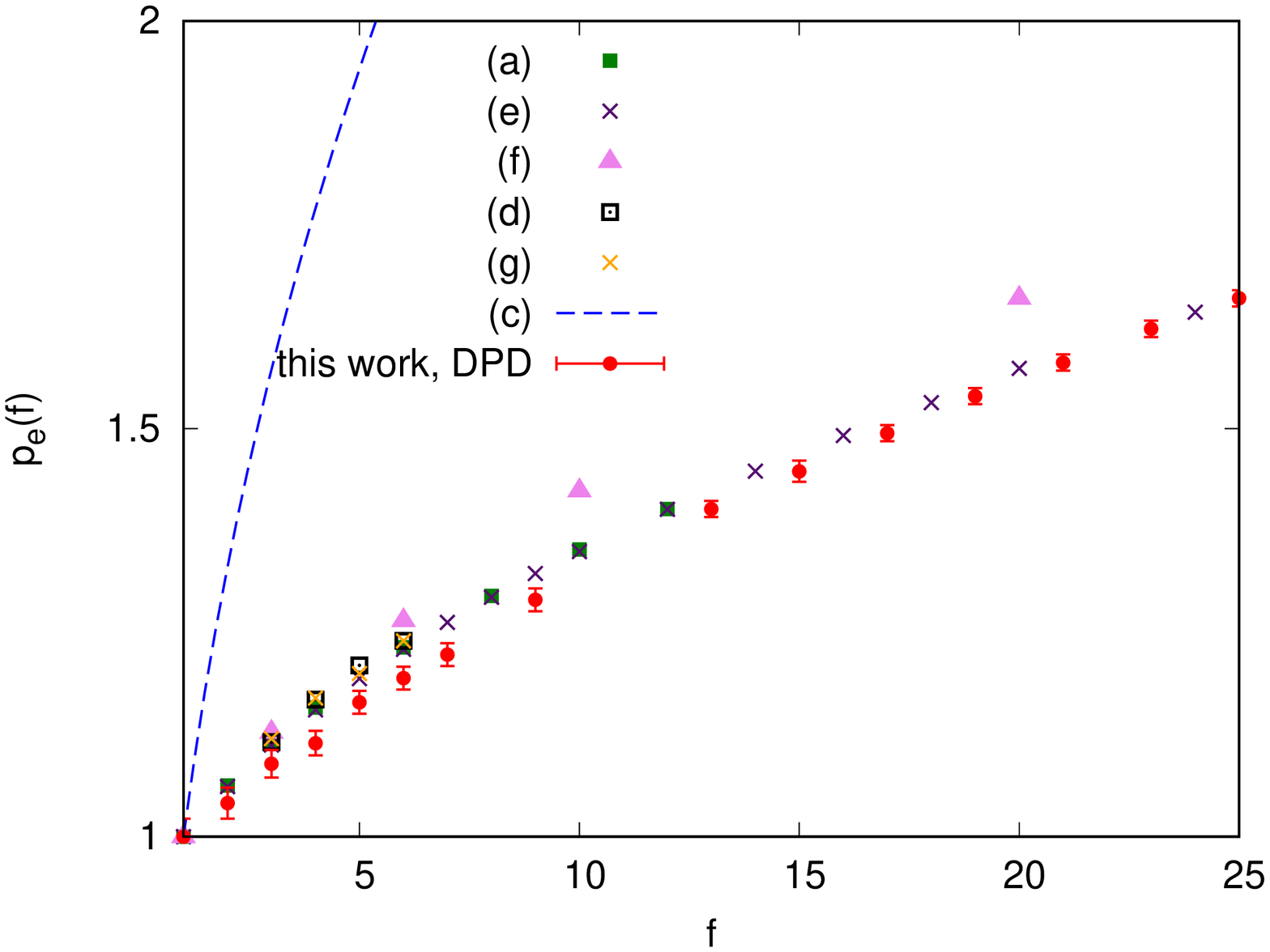}\\
\includegraphics[width=0.5\textwidth,angle=270]{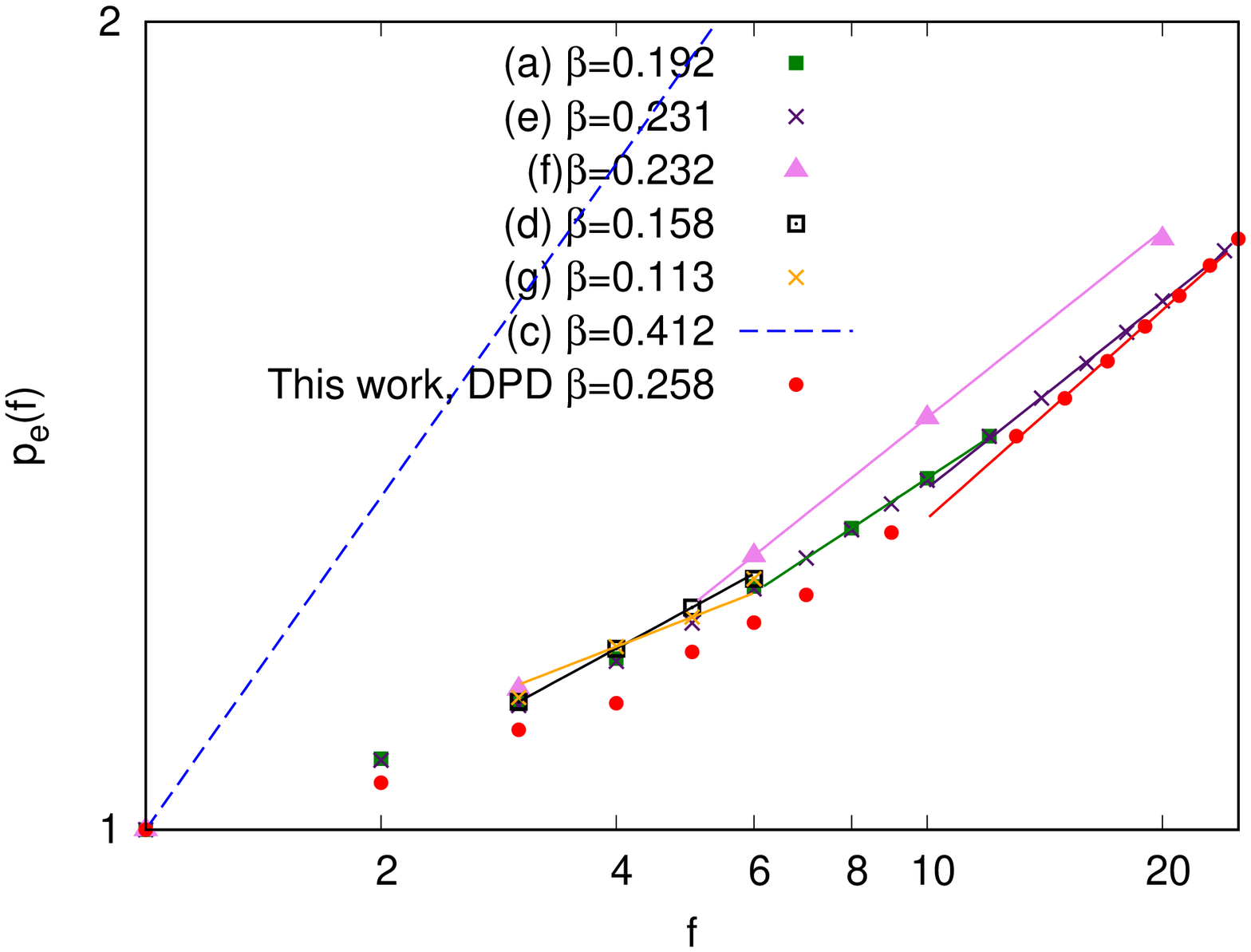}
\caption{\label{SAW_pe-arms} The same as in Fig.~\ref{SAW_g-star} but for the arm stretch ratio $p_e$ (\ref{pef_pgf}) and for the related exponent $\beta$ in Eq.~(\ref{beta}).}
\end{center}
\end{figure}

Now we will proceed to the shape characteristics for a single arm within a star polymer with $f$ branches. The arm stretch ratio $p_e$ (\ref{pef_pgf}) increases with the increase of $f$ due to the excluded volume effect of the neighboring branches, as shown in Fig.~\ref{SAW_pe-arms}. Comparison is performed against other simulation studies. The earlier simulations \cite{Whittington1986,Batoulis1989} were limited to the maximum values of $f=6$ or has a limited set of $f$ values \cite{Grest1994}, therefore the comparison of our data is focused on the results from Refs.~\cite{Zifferer1999} and \cite{Hsu2004}. We should note that, in general, our results for $p_e$ are close but found somehow lower than the data from both latter studies, especially for $f<10$. We may speculate that this might be the result of the soft interaction potentials used in the DPD simulations. Similarly to the cases of $g(f)$ and $p_A(f)$, we observe that the logarithmic scale plots in Fig.~\ref{SAW_pe-arms} are getting linear at larger $f$ values and one may write
\begin{equation}\label{beta}
p_e(f) \sim E f^\beta,~~~f \gg 1,~~N=\mathrm{const}.
\end{equation}
The value of $\beta=0.258$ is obtained in this study and, as follows from the comparison with the other simulation studies made in Fig.~\ref{SAW_pe-arms}, is relatively close to $\approx 0.23$ obtained by fitting the data from Refs.~\cite{Hsu2004} and \cite{Grest1994}. One should note that in all these cases the star polymers with at least $f=20$ branches have been analyzed. Lower values of $\beta$ in other simulations must be attributed to the fact that at the maximum values of $f$ considered there, the asymptotic behavior is not yet reached. We will also note that, in contrast to the case of $g(f)$, the Daoud-Cotton ansatz leads to the $p_e(f)$ dependence which is far away from all the simulations data. The value $\beta=1-3\nu\approx 0.412$ provided by this ansatz (see, Eq.~(\ref{g_pe_Daoud})) is an essential overestimate, as already been discussed earlier \cite{Zifferer1999,Havrnkova03,Hsu2004}.

\begin{figure}[t!]
\begin{center}
\includegraphics[width=0.5\textwidth,angle=270]{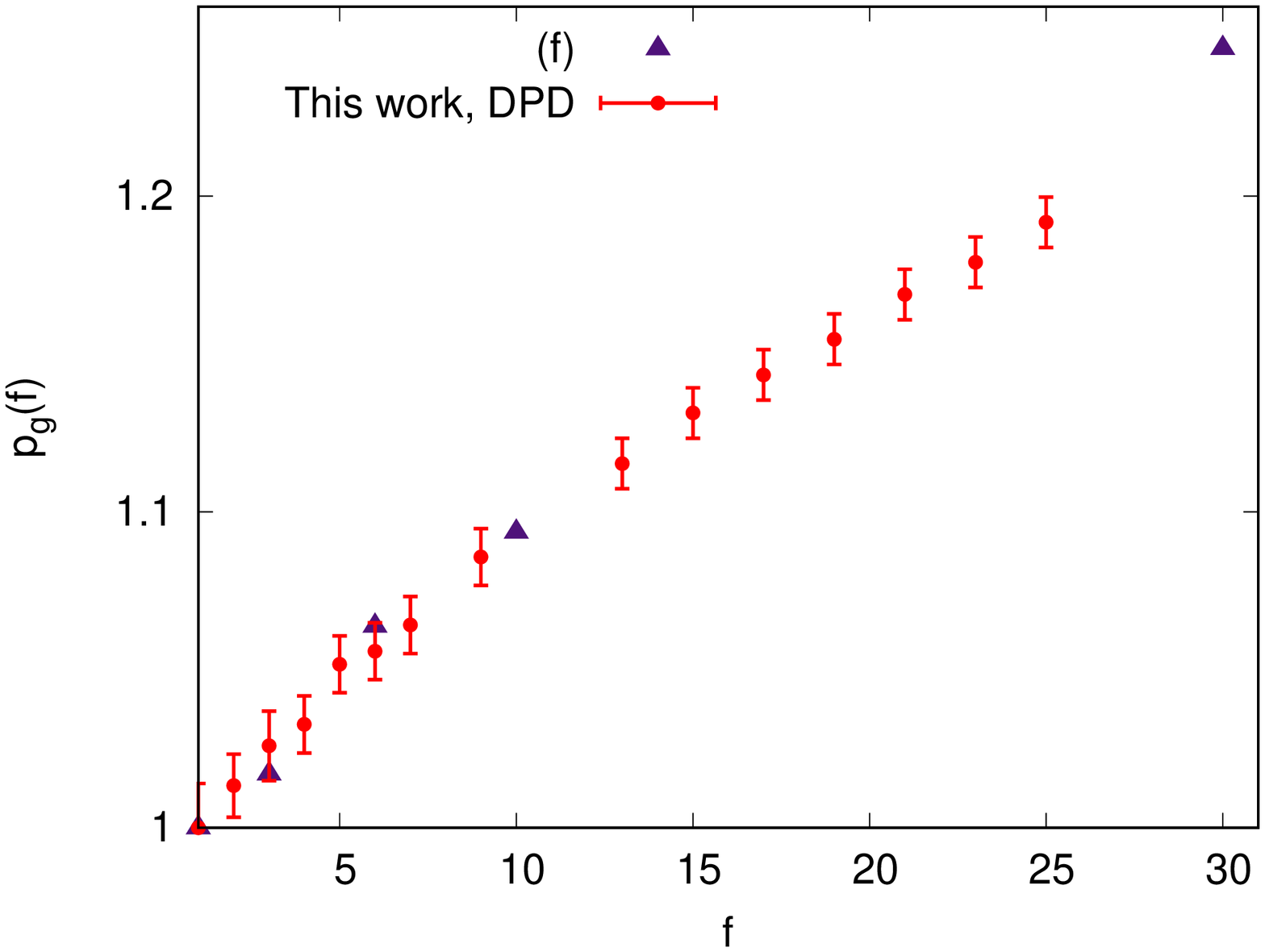}\\
\includegraphics[width=0.5\textwidth,angle=270]{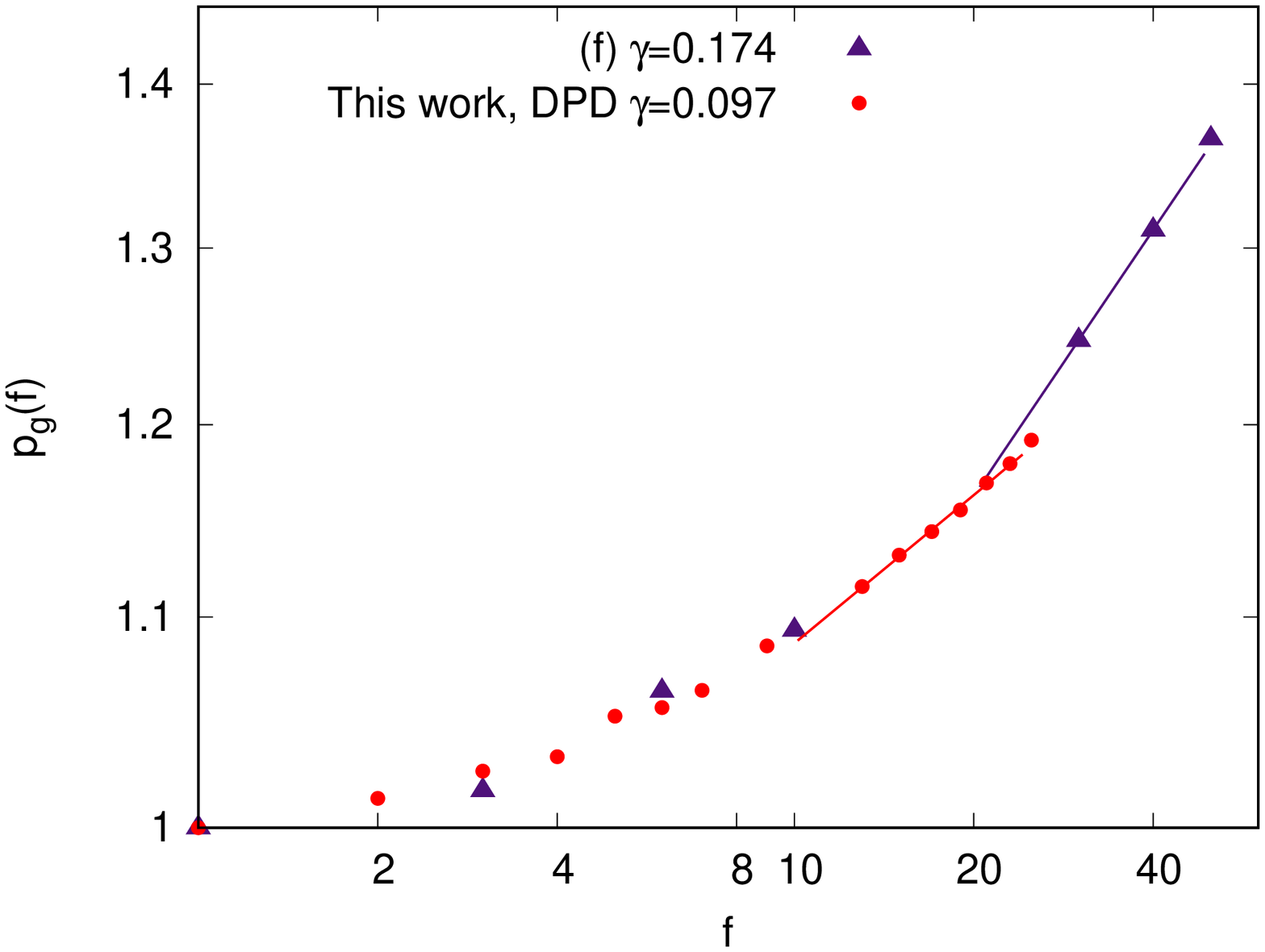}
\caption{\label{SAW_pg-arms} The same as in Fig.~\ref{SAW_g-star} but for the arm swelling $p_g$ ratio (\ref{pef_pgf}) and for the related exponent $\gamma$ in Eq.~(\ref{gamma}).}
\end{center}
\end{figure}
The arm swelling ratio, $p_g$, defined in Eq.~(\ref{pef_pgf}), characterizes the level of ``swelling'' of an individual arm due to the presence of other branches as seen via the increase of its squared gyration radius compared to that of the equivalent length freely suspended chain. Somehow it has got considerably less attention in the previous simulation studies. We show $p_g(f)$ evaluated in this study, alongside with the other available data found in Ref.~\cite{Grest1994}. We observe that our DPD results underestimate the value of this property as compared to the molecular dynamics study of Ref.~\cite{Grest1994}, especially at $f>10$. This is reflected in the lower value of the exponent $\gamma\approx0.097$ when fitting our data to the power law
\begin{equation}\label{gamma}
p_g(f) \sim F f^\gamma,~~~f \gg 1,~~N=\mathrm{const}.
\end{equation}
as compared to $\gamma\approx0.145$ from the analysis of the Ref.~\cite{Grest1994} data. For the behavior of $p_e(f)$ we observed that the DPD results underestimated its value at smaller $f<10$ as compared to the molecular dynamics studies but provided the same scaling behavior in terms of the exponent $\beta$. In contrary, for the $p_g(f)$, the difference is found both at smaller $f$ and for its scaling behavior of $p_g(f)$. However, to made some definite statement, more comparison between various simulations is
needed.

\begin{figure}[t!]
\begin{center}
\includegraphics[width=0.5\textwidth,angle=270]{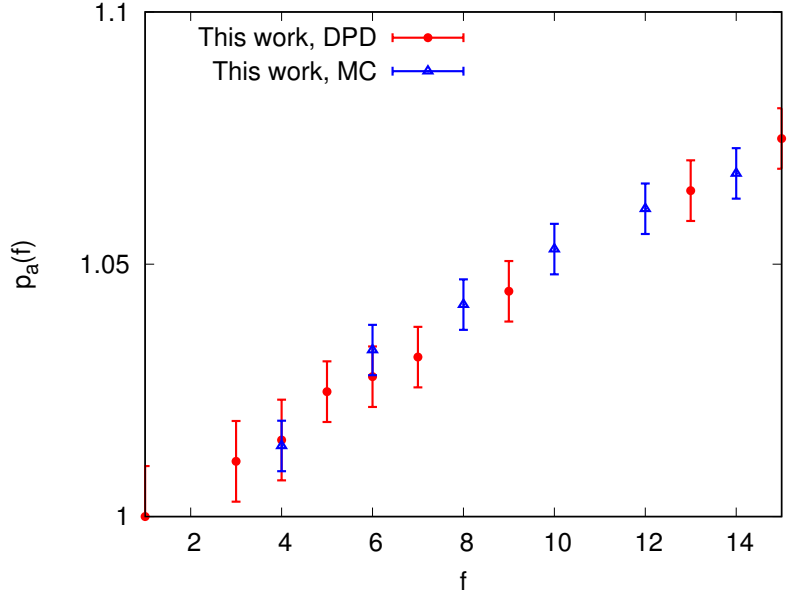}\\
\caption{\label{SAW_pa-arms} The arm asphericity ratio $p_a$ (\ref{paaf}), comparison between the dissipative particle dynamics and Monte Carlo simulation results, both obtained within this study.}
\end{center}
\end{figure}

Finally, we analyze the level of asphericity of individual arm in the presence of other branches as given by the $p_a$ ratio defined in Eq.~(\ref{paaf}).
Since no other numerical studies of  this quantity are available so far, for comparison we performed also lattice simulations based on the Monte Carlo growth chain algorithm
in addition to dissipative dynamics simulation.
Results are shown in Fig.~\ref{SAW_pa-arms} as a function of the number of branches $f$. As expected, we observe an increase of $p_a$ with growing $f$:
effect of crowdedness, caused by presence of our branches, leads to straightening of individual arm, how ever the effect is rather weak with $p_a(f)$ not exceeding $1.08$ for $f=15$

\section{\label{IV}Conclusions}

We analyzed the set of properties, which allow to characterize the impact of
local crowdedness caused by structure of $f$-branched star polymer on the peculiarities of spatial extension of single arm.
To this end, we consider the characteristics, specific to an individual arm within  the star, such as the average center-end distance
$\langle r^2_{e,1} \rangle$,  the average squared gyration radius
$\langle r^2_{g,1} \rangle$ and the asphericity of an individual arm within a star $a_{f}$.
The corresponding universal ratios $p_e(f)$, $p_g(f)$ and $p_a(f)$, as given by Eqs. (\ref{pef_pgf}) and (\ref{paaf}),
are introduced, to compare these values directly with that of a freely suspended linear chain of the same molecular weight.
Note these values  are independent on any details of chemical structure of molecules and are governed only by global parameters, such as dimension
of space $d$ or functionality $f$.

The main theoretical findings of this study are collected in Eqs.~(\ref{eEV}), (\ref{gEV}) and (\ref{aEV}). As follows from these expressions, all three arm shape ratios increase linearly with the number of chains $f$. The simulation results for these characteristics are given in Figs.~\ref{SAW_pe-arms}, \ref{SAW_pg-arms} and \ref{SAW_pa-arms} and they show the same trend but display the saturation-like behavior.

When comparing theoretical findings and the simulation data side-by-side, one should make several remarks. The first one is that the theoretical model being used here assumes the limit of an infinite length of each arm. In this case, the infinite values for $p_e(f)$, $p_g(f)$ and $p_a(f)$ at $f\to\infty$ can be considered as, at least, feasible. In contrast to that, any simulation model deal with the arms of finite length, therefore, there is a limit for $p_e(f)$, $p_g(f)$ and $p_a(f)$ at large $f$ which is always finite. For instance, for the case of $p_e(f)$, if we assume completely stretched conformation for an individual chain at $f\to\infty$, then $\langle r_{e,\infty}\rangle\sim (N-1)l$ ($N=8$ and $l\approx 0.9$ is an average bond length in the DPD simulations), whereas $\langle r_{e,1}\rangle\sim ((N-1)l)^{\nu}$, where $\nu=0.588$ assuming good solvent condition. This yields the maximum possible value of $p_e(\infty)\approx 2.13$, which looks not unreasonable given the shape of the $p_e(f)$ in Fig.~\ref{SAW_pe-arms}. Therefore, one source of discrepancies between the limit values at $f\to\infty$ is quite evident.

The second remark is that the theoretical calculations are done in the first order of $\epsilon=4-d$. It is known that for some properties, e.g. the critical exponents $\nu$ and $\gamma$, these terms provide the main contributions to the Gaussian results towards the values estimated by other means. However, this must be not necessarily true for the case of the properties of interest considered in this study. The answer can possibly be found only by undertaking the calculations to higher orders in $\epsilon$.

Another finding, that can be pointed out as the result of this study, is the fact of a really good agreement for a number of shape characteristics when the data from the DPD simulations are compared with the results of Monte Carlo and molecular dynamics. This is especially valuable, as far as the increase of all shape characteristics is due to excluded volume effect and crowding of star arms. DPD uses soft repulsive interaction potentials and there were doubts \cite{Kong1997,Spenley2000} whether such type of potential is able to describe the excluded volume effects adequately. Providing some answers in Refs.~\cite{Ilnytskyi2007,Kalyuzhnyi2016,Kalyuzhnyi2018}, here we also show that the method describes adequately the excluded volume effect in the case of dense star with a relatively large number of arms.

\section*{Acknowledgements}
The authors acknowledge allocation of computer time for the simulations at the cluster of ICMP of the National Academy of Science of Ukraine and Ukrainian National Grid.
\bibliography{Star}

\newpage
\tableofcontents
\newpage
{\bf Text for the Table of Contents}
\begin{figure}[t!]
\begin{center}
\includegraphics[width=0.19\textwidth,angle=200]{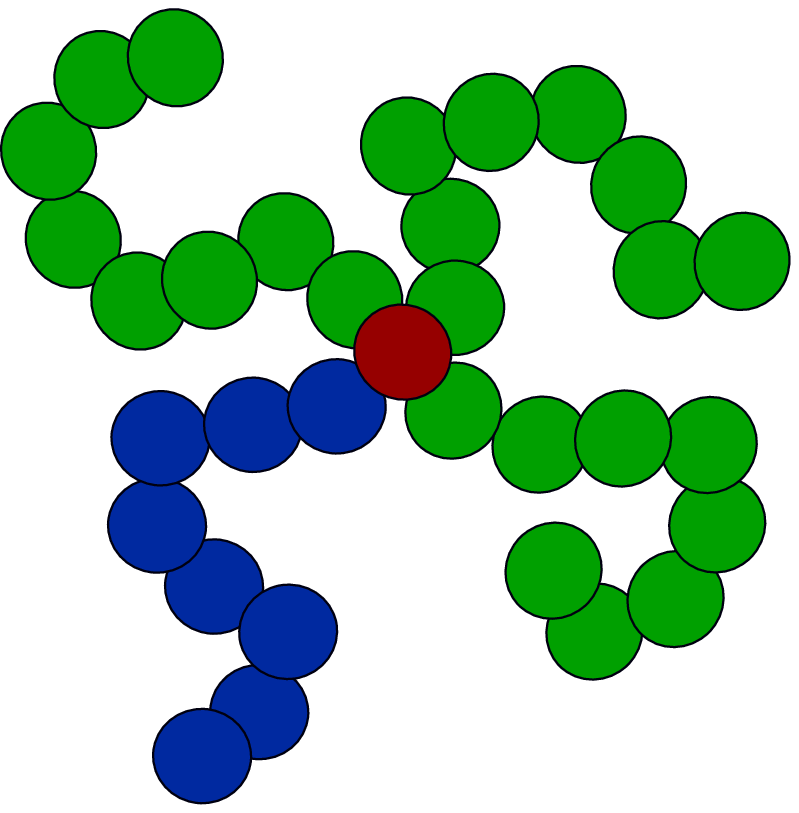}$\qquad\qquad\qquad$\includegraphics[width=0.3\textwidth,angle=270]{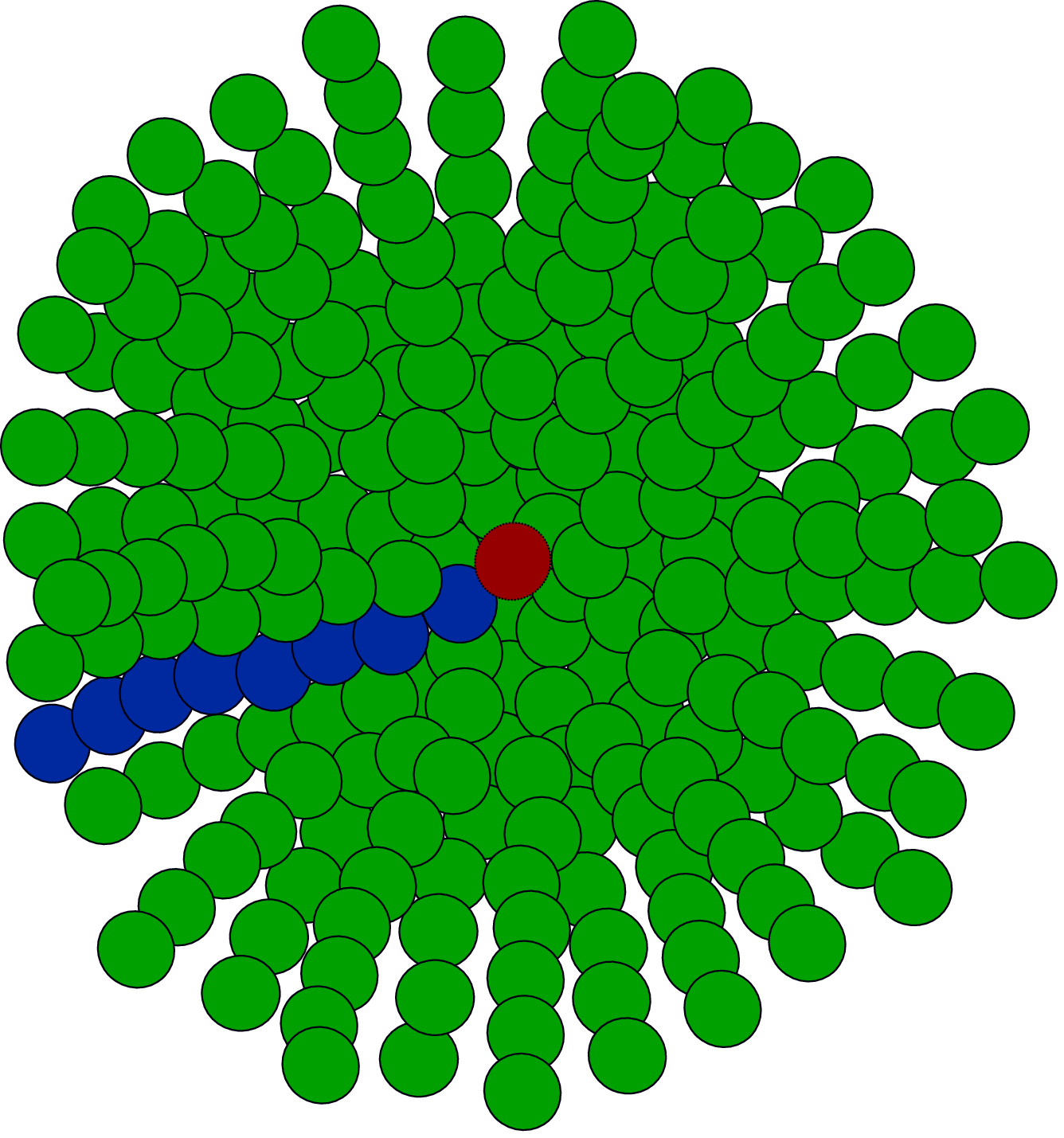}
\end{center}
\end{figure}

How does an arm of a star polymer stretch? We apply both numerical and analytical approaches to get an answer. Analytical description is given by direct polymer renormalization and a numerical one is achieved by DPD simulations and in parts by lattice MC method. The results are presented by a set of size ratios comparing properties of arm and a free polymer chain.

\end{document}